\documentclass[aps,twocolumn,showpacs,preprintnumbers,prl,superscriptaddress]{revtex4}

\usepackage{graphicx}% Include figure files
\usepackage[version=3]{mhchem} % Formula subscripts using \ce{}
\usepackage{amsmath}
\usepackage{amssymb}
\usepackage{graphicx}
\usepackage{cleveref}

\newcommand{\R}{\mathbf{r}}

\newcommand{\A}{\text{A}}
\newcommand{\B}{\text{B}}

\newcommand{\nadd}{\text{nadd}}
\newcommand{\GGA}{\text{GGA}}

\newcommand{\s}{\text{s}}
\newcommand{\xc}{\text{xc}}
\newcommand{\xonly}{\text{x}}
\newcommand{\conly}{\text{c}}
\newcommand{\KS}{\text{KS}}
\newcommand{\abs}[1]{\left|#1\right|}

\def\gks{\text{GKS}}

\begin{document}

\title{Semilocal and Hybrid Density Embedding Calculations of Ground-State Charge-Transfer Complexes}

\author{S. Laricchia}
\affiliation{Center for Biomolecular Nanotechnologies @UNILE, Istituto Italiano di Tecnologia (IIT), 
Via Barsanti, 73010 Arnesano (LE), Italy}
\author{E. Fabiano}
\affiliation{National Nanotechnology Laboratory (NNL), Istituto Nanoscienze-CNR,
Via per Arnesano 16, 73100 Lecce, Italy}
\author{F. Della Sala}
\affiliation{National Nanotechnology Laboratory (NNL), Istituto Nanoscienze-CNR,
Via per Arnesano 16, 73100 Lecce, Italy}
\affiliation{Center for Biomolecular Nanotechnologies @UNILE, Istituto Italiano di Tecnologia (IIT), 
Via Barsanti, 73010 Arnesano (LE), Italy}

\date{\today}

\begin{abstract}
We apply the frozen density embedding method, using a full relaxation of
embedded densities through a freeze-and-thaw procedure, to study the
electronic structure of several benchmark ground-state charge-transfer
complexes, in order to assess the merits and limitations of the approach for
this class of systems. The calculations are performed using both semilocal and
hybrid exchange-correlation (XC) functionals.
The results show that embedding calculations using semilocal XC functionals
yield rather large deviations with respect to the corresponding supermolecular
calculations. Due to a large error cancellation effect, however, they can
often provide a relatively good description of the electronic structure of
charge-transfer complexes, in contrast to supermolecular calculations
performed at the same level of theory. On the contrary, when hybrid XC
functionals are employed, both embedding and supermolecular calculations agree
very well with each other and with the reference benchmark results.

In conclusion, for the study of ground-state charge-transfer complexes via
embedding calculations hybrid XC functionals 
are the method of choice due to their higher reliability and 
superior performance.
\end{abstract}

\maketitle

\section{Introduction}
\label{sec:intro}
In the last years a considerable interest in density-embedding methods has
led to a widespread development of subsystem approaches \cite{yang91,yang95,
koba07,he09,lwwang08,lwwang08b,sekino03,fedorov04dft,senatore,cortona,
cortona94,cohen07_2,cohen07,huang06,huang11,cohen09,elliot09,
nafziger11,gordon12,gomes12,pavanello2}
within density functional theory (DFT) \cite{hohenberg-kohn,KS}. 
In particular, the frozen density embedding (FDE) method \cite{wesowarh93,
wesobook,dulak07,kevorkyants06hyd,wesolowski97hyd,WVW98,
tranall02,neugebauer05solv,jacob08,Neugebauer05,jacob06solv,
milleriii12,pavanello1}
has emerged as a popular tool to treat intermolecular interactions in
a simple and potentially exact embedding scheme.

In the FDE approach the total electronic density of a system is partitioned
into a set of subsystem densities (e.g., for two subsystems $\A$ and $\B$: 
$\rho=\rho_\A+\rho_\B$). The total electronic energy is then written as
\begin{eqnarray}
\nonumber
&&E[\rho] = E[\rho_\A+\rho_\B] = T_\s[\rho_\A] + T_\s[\rho_\B] + 
T_\s^{\nadd}[\rho_\A,\rho_\B] + \\
\nonumber
&& + \frac{1}{2}\int\int\frac{\left(\rho_\A(\R_1)+\rho_\B(\R_1)\right) 
\left(\rho_\A(\R_2)+\rho_\B(\R_2)\right)}{\abs{\R_1-\R_2}}\,d\R_1d\R_2 + \\
\nonumber
&& + \int \left(v_{\text{ext}}^\A(\R_1)+v_{\text{ext}}^\B(\R_1)\right)
\left(\rho_\A(\R_1)+\rho_\B(\R_1)\right)d\R_1 + \\
\label{e1}
&&+ E_\xc[\rho_\A] + E_\xc[\rho_\B] + E_\xc^\nadd[\rho_\A,\rho_\B]\ ,
\end{eqnarray}
where $T_\s$ is the noninteracting kinetic energy functional, the
external potential was partitioned as $v_{\text{ext}}=v_{\text{ext}}^\A
+v_{\text{ext}}^\B$, and 
the superscript $\nadd$ denotes the nonadditive energy contributions to 
the kinetic and exchange-correlation (XC) energies.
The minimization of the energy of Eq. (\ref{e1}) with respect to
one of the subsystem electron densities (e.g., $\rho_\A$), keeping
the other one fixed and using the expansion of the electron density in terms of
(auxiliary) Kohn-Sham orbitals $\psi_i^\A$, leads to the Kohn-Sham equations with 
constrained electron density (KSCED) \cite{wesowarh93,wesobook}
\begin{equation}\label{e2}
\left(-\frac{1}{2}\nabla^2 + v^\KS[\rho_\A](\R) + 
v_{\text{emb}}[\rho_\A;\rho_\B](\R)\right)\psi^\A_i = \varepsilon_i^\A\psi_i^\A
\end{equation}
with 
\begin{eqnarray}
\nonumber
v^\KS[\rho_\A](\R) & = & \int\frac{\rho_\A(\R_1)}{\abs{\R-\R_1}}\,d\R_1 + 
v_{\text{ext}}^\A(\R) + \frac{\delta E_\xc[\rho_\A]}{\delta\rho_\A(\R)} \\
\label{e3_vks}
&& \\
\nonumber
v_{\text{emb}}[\rho_\A;\rho_\B](\R) & = & v_{\text{ext}}^\B(\R) + 
\int\frac{\rho_\B(\R_1)}{\abs{\R-\R_1}}\,d\R_1 + \\
\label{e3}
&+& \frac{\delta T_\s^{\nadd}[\rho_\A,\rho_\B]}
{\delta\rho_\A(\R)} + \frac{\delta E_\xc^{\nadd}[\rho_\A,\rho_\B]}
{\delta\rho_\A(\R)} \ . 
\end{eqnarray}
The KSCED of Eq. (\ref{e2}) provide the solution for 
the ground state of subsystem
$\A$ subject to the interaction with the frozen density of the subsystem $\B$.
If the KSCED of both subsystems are considered in a freeze-and-thaw procedure
\cite{weso96iter,wesobook}, the full variational solution for the total system, 
equivalent to the usual Kohn-Sham solution, is obtained, except for 
approximations eventually included in the nonadditive interaction terms.
Henceforth, the acronym FDE wil be used to refer to this approach.

So far, numerous applications of the FDE-based 
method, using the KSCED approach, have 
been presented in literature to study the interaction of 
non-covalently bound (or weakly interacting) molecular complexes
or solvated molecules \cite{dulak07,kevorkyants06hyd,wesolowski97hyd,WVW98,
tranall02,neugebauer05solv,jacob08,Neugebauer05,jacob06solv}.
In this cases in fact the nonadditive kinetic energy term can be efficiently 
approximated by means of semilocal approximations \cite{lembarki94,
constantin2011,laricchia2011,beyhan10}. 
However, because in the conventional formulation the use of
KSCED is limited to local/semilocal XC functionals,
most studies have focused on hydrogen-bond 
and dipole-dipole systems, since these interactions are  
dominated by electrostatic and polarization effects that can
be described relatively well at the semilocal level of theory
\cite{2truhlar05,mukappa}.

Recently the method has been also extended to the generalized
Kohn-Sham (GKS) framework \cite{laricchia2010,larihyben}, 
so that hybrid functionals can be considered
as well. In this approach, named GKS-FDE, the 
nonadditive kinetic and XC terms in Eq. (\ref{e3})
are replaced by the nonadditive direct and residual interactions $F_\s^{\nadd}$
and $R_\s^{\nadd}$ which include respectively the orbital-dependent and the 
explicitly density-dependent kinetic and XC terms.
The presence of nonlocal and/or orbital-dependent terms makes 
however the direct implementation of the method rather involved. 
Therefore, a practical computational scheme was 
proposed where these terms are approximated with 
local functionals \cite{laricchia2010}.
Thus, in practice in the GKS-FDE method the electronic 
structure is evaluated using a 
nonlocal exchange operator for the intra-subsystem XC effects but with
an embedding potential evaluated at the local/semilocal level of theory.
A similar approximation was also proposed for the extention of the
KSCED to the use of orbital-dependent optimized 
effective potentials \cite{laricchia-lhf2011}.

The GKS-FDE method was tested for a variety of hydrogen bond and 
dipole-dipole interactions and proved to be a very good approach for 
embedded simulations with hybrid 
functionals \cite{laricchia2010,laricchia-lhf2011,larihyben}.
In particular, with respect to FDE-based calculations employing semilocal
XC functionals, it enables a general reduction of the embedding errors,
because of the smaller overlap of subsystem densities at the hybrid 
level, and it allows a better description of subsystem properties
due to the removal of most of the self-interaction error (SIE)
\cite{cohen-sie,dellasala-sie,laricchia-lhf2011}.
Furthermore, it makes it possible to extend the application of FDE-based 
techniques to treat systems that are typically poorly described by
semilocal XC functionals.

An example of the latter is given by ground-state charge-transfer complexes
(also called donor-acceptor complexes). These systems derive in fact a 
considerable portion of their stabilization energy from an electron 
charge-transfer between two centers. Thus, unlike for hydrogen bonds
or dipole-dipole interactions, where electrostatic and polarization
effects dominate, the interaction in charge-transfer complexes is
largely determined by orbital interactions. As a consequence, the
description of ground-state charge-transfer complexes is very
challenging for conventional semilocal XC functionals because of the SIE
inherent in these methods. On the other hand, a good performance is
generally obtained for hybrid functionals, especially when rather large 
portions of Hartree-Fock exchange are included \cite{steinmann12,
ruiz95,ruiz96,garcia97,garcia2000,karpfen03,liao_2-03,liao03,smith10,
timo08,phil07,sze02,li02,stein09,manna09,bhatta07,lopez09}. Similar results
also concern the charge transfer between organic molecules and
metal substrates \cite{JCPacc,dellasala10}.

Due to the difficult application of DFT-based methods using semilocal XC
functionals to these systems, in the last years only few FDE investigations
treated ground-state charge-transfer complexes and related problems
\cite{gotz09,beyhan10,dulak06,garcialastra08,pavanello1}.
Some of them \cite{beyhan10,garcialastra08} made use of complicated or
{\it ad hoc} reparametrized kinetic energy functionals in order 
to obtain improved kinetic embedding potentials and improve the
description of the system. However, these kinetic functionals cannot be
routinely applied in FDE calculations for large systems.
On the other hand, an extensive investigation of different kinetic energy
functionals \cite{gotz09} showed that most popular kinetic 
energy approximations perform similarly and reasonably well for 
complexes displaying small/moderate charge transfer, so that 
it appears reasonable to consider in such applications the same 
kinetic energy approximation as those commonly used for hydrogen bond
and dipole-dipole interactions.
The role of the XC functionals in embedding calculations dealing
with charge-transfer systems, instead, was not given much
attention so far \cite{pavanello1}.

In this paper we cover this issue and
demonstrate how hybrid functionals can be practically utilized
within an embedding scheme to obtain an improvement of the performance in the 
description of embedded systems for a benchmark of charge-transfer complexes.

\section{Method and computational details}
\label{sec:comdet}

The KSCED \cite{wesowarh93,wesobook} and 
GKS-FDE \cite{laricchia2010,larihyben} 
approaches are implemented in the TURBOMOLE program package 
\cite{TURBOMOLE}, version 6.4.
All calculations were performed using the {\it FDE} script, 
which uses a freeze-and-thaw procedure 
\cite{weso96iter} to guarantee the full relaxation of the 
embedded ground-state electron density of both subsystems, 
until dipole moments of the embedded subsystems converged to $10^{-3}$ au. 
A supermolecular def2-TZVPPD \cite{weigend03,weigend05} 
basis set was employed in all calculations to expand the subsystem 
electron densities. 
A monomolecular (or too small) basis set is in fact insufficient
to carry on FDE calculations with good accuracy, resulting in 
basis-set errors larger than the FDE errors and the differences
between the methods.
The def2-TZVPPD basis set adds diffuse basis functions to 
the def2-TZVPP \cite{weigend05} basis set, thus granting an accurate 
representation for the electron densities even at the relatively 
large bonding distances characteristic of the systems under consideration.
Very accurate integration grids were employed to minimize numerical errors.
Additional details about our implementation and computational procedure 
are reported in Refs. ~\citenum{laricchia2010,larihyben,laricchia-lhf2011}.

In our investigation we considered a set of twelve representative 
ground-state charge-transfer complexes (C$_2$H$_4$-F$_2$, NH$_3$-F$_2$, 
C$_2$H$_2$-ClF, HCN-ClF, NH$_3$-Cl$_2$, H$_2$O-ClF, NH$_3$-ClF, HCN-NF$_3$, 
HNC-NF$_3$, HF-NF$_3$, ClF-NF$_3$, and C$_2$H$_4$-Cl$_2$).
The geometries and the reference binding energies were taken from
Refs. ~\citenum{truhlar05,steinmann12}, except for the
C$_2$H$_4$-Cl$_2$ complex (for this latter case we optimized the 
geometry at the MP2/aug-cc-pVTZ level of theory \cite{mp2,dunning89,augdunning,
delbene93} and calculated the
reference binding energy, in analogy with Ref. \cite{pitonak08}
as $E_{b}^{ref}=E_b^{\text{CCSD(T)}} + \Delta E_b^{\text{MP2}}$,
where $E_b^{\text{CCSD(T)}}$ is the binding energy computed at the 
CCSD(T)/aug-cc-pVQZ level of theory \cite{ccsd_t,dunning89,augdunning,
delbene93} and $\Delta E_b^{\text{MP2}}$
is the difference between the MP2 binding energies computed at the 
complete basis set limit (56-extrapolation \cite{halkier98})
and with an aug-cc-pVQZ basis set; the resulting binding
energy is in very good agreement to that reported in Ref. ~\citenum{ruiz96}).

The calculations were performed using conventional semilocal XC functionals 
(BLYP \cite{BEC88,lyp}, PBE \cite{pbe}) as well as hybrid methods 
(PBE0 \cite{pbe0-perdew} and BHLYP \cite{BEC88,lyp}).
All calculations were performed 
using DFT-D3 dispersion corrections \cite{disp3}.  
To compute the embedding nonadditive interactions, 
the nonadditive noninteracting kinetic energy term was approximated 
using the GGA functional revAPBEk \cite{constantin2011,laricchia2011}.
In the case of hybrid XC functionals also the nonadditive exchange term
was approximated: the PBE and the B88x \cite{BEC88} exchange
functionals were employed for PBE0 and BHYLP, respectively;
the resulting methods are indicated as PBE0/PBE and BHLYP/BLYP.
In the case of semilocal XC functionals instead, no approximation for the 
nonadditive exchange contribution to the embedding potential is required:
the same exchange functional is used for the embedding potential, subsystem 
and conventional supermolecule KS calculations.

The quality of embedded densities was investigated by considering the absolute 
deviation of plane-averaged embedding density, defined as
\begin{equation}\label{drho}
\Delta\bar{\rho}(z) = \int \abs{{\rho}_{\A}(x,y,z)+{\rho}_{\B}(x,y,z) - 
\rho^{\gks}(x,y,z)}\,dxdy \,
\end{equation}
where $\rho^{\gks}$ is the electron density
obtained from a (generalized) KS calculation on the whole system.
A quantitative measurement of the absolute error associated with a given
embedded density was then obtained by computing the embedding density error
\begin{equation}\label{eq:xi}
\xi = \frac{1000}{N} \int \Delta\bar{\rho}(z)dz
\end{equation}
with $N$ being the number of electrons.
In the evaluation of $\xi$ only valence densities were considered to avoid
numerical problems related to the high value assumed by the electron density
in the core regions and in consideration of the fact that the core density 
is not important for the determination of chemical and physical properties 
of interaction between the subsystems, which are of interest here. 

The net charge on the $i$-th subsystem (monomer charge)
was calculated for each method by considering a grid-based 
atoms-in-the-molecule partitioning \cite{aim} of the electron density into 
atomic basins $\Omega_j$ as
\begin{equation}\label{eq:ct}
\nu_i = \sum_{j=1}^{N_i}\int_{\Omega_j}\rho(\R)d\R\ - \ Z_j 
\end{equation}
with $Z_j$ the nuclear charge of the $j$-th atom and $N_i$ the number of atoms 
in the $i$-th subsystem. 
For two interacting subsystems $\nu_\A+\nu_\B=0$ and thus we define
the charge transfer as $\chi=\abs{\nu_\A}=\abs{\nu_\B}$.
Reference values for the charge transfer were 
evaluated using relaxed MP2 densities computed with an aug-cc-pVTZ basis set.
 
Finally, to inspect the quality of the embedding energies we considered the
error defined as
\begin{equation}\label{eq_emberr}
\Delta E[\rho_\A,\rho_\B,\rho^{\gks}] = E^{\text{emb}}[\rho_\A,\rho_\B] 
- E^{\gks}[\rho^{\gks}]\ ,
\end{equation}
where $E^{\text{emb}}$ is the embedding energy, computed using 
the embedding densities
into Eq. (\ref{e1}), and $E^{\gks}$ is the energy resulting from a conventional
(generalized) KS calculation on the full system.
Further analysis was also performed considering, according 
to Ref. ~\citenum{larihyben}, the error decomposition
\begin{eqnarray}
\nonumber
\Delta E[\rho_A,\rho_B,\rho^\gks] & = & \Delta T[\rho_A,\rho_B,\rho^\gks]+ 
\Delta D[\rho,\rho^\gks] +\\
\label{eq:enerror}
&& + \Delta X^S[\rho_A,\rho_B,\rho^\gks]\ ,
\end{eqnarray}
where the error on embedding energy is partitioned into
kinetic ($\Delta T$), exchange ($\Delta X^S$), and relaxation
($\Delta D$) contributions. For full details see Ref. ~\citenum{larihyben}.

\section{Results and discussion}
\label{sec:results}

In this section we report and analyze the results of the FDE-based
calculations on the charge-transfer complexes. 
We evaluate the performance of hybrid and
semilocal XC functionals (BHLYP, PBE0, BLYP, and PBE),
elucidating what are the roles of the exact exchange, of the geometry and
of the charge transfer between the subsystems.

\subsection{Embedding energy}
\label{sec:enerr}

In Tab. \ref{tab:enerr} we report the binding energies obtained from
different embedding methods ($E_b^{\text{emb}}$) and the corresponding results from 
conventional DFT calculations on the whole supermolecular system ($E_b^\gks$).
In addition, we report the deviation of these results from accurate reference 
values ($\Delta^{\text{emb}}$ and $\Delta^\gks$, respectively) and the error on 
embedding energies as defined in Eq. (\ref{eq_emberr}). 
We note that the embedding error 
on the total energy is exactly equivalent to the difference of the 
binding energies computed respectively from embedding and conventional
supermolecular calculations (i.e. the DFT calculation on the whole system), 
since the contributions from isolated subsystems exactly cancel out. 
Moreover, all binding energies reported in Tab. \ref{tab:enerr} are
dispersion-corrected and, because we use a supermolecular basis set, the
basis set superposition error for the total-system calculation is treated in
the same way in both DFT and FDE-based methods (Boys-Bernardi correction
\cite{bsse2}).
In Tab. \ref{tab:enerr} we also report in the last lines
the mean signed error (MSE), the mean absolute error (MAE), 
and the mean absolute relative error (MARE)
with respect to the reference binding energies.
\begin{table*}
\caption{\label{tab:enerr} Dispersion-corrected binding 
energies resulting from conventional 
supermolecular DFT ($E_b^\gks$) and embedding ($E_b^{\text{emb}}$) calculations 
for several test charge-transfer complexes. The differences of the binding 
energies with respect to the reference ones are also reported 
($\Delta^\gks=E_b^\gks-E_b^{ref}$ and $\Delta^{\text{emb}}=
E_b^{\text{emb}}-E_b^{ref}$) as well as the error 
on the embedding energy ($\Delta E$; 
see Eq. (\ref{eq_emberr})). The reference binding energies ($E_b^{ref}$) 
are reported in the second column. 
In all calculations we used the revAPBEk functional for the non-additive
kinetic energies and a supermolecular def2-TZVPPD basis set.
In the last lines we report the mean 
absolute error (MAE), the mean signed error (MSE), and the mean absolute 
relative error (MARE). All values are in mHa.}  
%\begin{scriptsize}
\begin{tabular}{lrcrrrrrcrrrrr}
\hline\hline
System & $E_b^{ref}$ & $\;\;$ & $E_b^\gks$ & $E_b^{\text{emb}}$ & $\Delta^\gks$ & 
$\Delta^{\text{emb}}$ & $\Delta E$ & $\;\;$ & $E_b^\gks$ & $E_b^{\text{emb}}$ & 
$\Delta^\gks$ & $\Delta^{\text{emb}}$ & $\Delta E$ \\ 
\hline
       &            &        & \multicolumn{5}{c}{BLYP} &        & \multicolumn{5}{c}{BHLYP} \\
\cline{4-8}\cline{10-14}
NF$_3$-HCN       & 1.67$^a$ & &  0.96 &  1.40 & -0.71 & -0.27 &  0.44 & & 1.71  & 1.11 & 0.04 & -0.56 & -0.60 \\
C$_2$H$_4$-F$_2$ & 1.69$^b$ & &  5.21 &  0.56 &  3.52 & -1.13 & -4.65 & & 2.24  & 0.11 & 0.55 & -1.58 & -2.13 \\
NF$_3$-HNC       & 2.31$^a$ & &  2.93 &  3.07 &  0.62 &  0.76 &  0.14 & & 2.56  & 2.35 & 0.25 &  0.04 & -0.21 \\
C$_2$H$_4$-Cl$_2$& 2.60$^c$ & &  6.25 &  4.76 &  3.65 &  2.16 & -1.49 & & 3.81  & 3.91 & 1.21 &  1.31 &  0.10 \\ 
NH$_3$-F$_2$     & 2.88$^b$ & &  8.22 &  1.16 &  5.34 & -1.72 & -7.06 & & 3.51  & 0.99 & 0.63 & -1.89 & -2.52 \\
NF$_3$-ClF       & 2.92$^a$ & &  4.14 &  2.14 &  1.22 & -0.78 & -2.00 & & 2.89  & 1.40 &-0.03 & -1.52 & -1.49 \\
NF$_3$-HF        & 2.92$^a$ & &  4.14 &  3.22 &  1.22 &  0.30 & -0.92 & & 3.52  & 2.71 & 0.60 & -0.21 & -0.81 \\
C$_2$H$_2$-ClF   & 6.07$^b$ & &  8.74 &  5.09 &  2.67 & -0.98 & -3.65 & & 6.96  & 5.26 & 0.89 & -0.81 & -1.70 \\
HCN-ClF          & 7.74$^b$ & &  8.17 &  6.70 &  0.43 & -1.04 & -1.47 & & 7.90  & 7.27 & 0.16 & -0.47 & -0.63 \\
NH$_3$-Cl$_2$    & 7.78$^b$ & & 11.82 &  9.23 &  4.04 &  1.45 & -2.59 & & 8.59  & 8.67 & 0.81 &  0.89 &  0.08 \\ 
H$_2$O-ClF       & 8.54$^b$ & & 10.30 &  8.11 &  1.76 & -0.43 & -2.19 & & 9.60  & 8.77 & 1.06 &  0.23 & -0.83 \\
NH$_3$-ClF       & 16.92$^b$& & 25.13 & 21.29 &  8.21 &  4.37 & -3.84 & & 21.03 & 20.84& 4.11 &  3.92 & -0.19 \\
%                 &          & &       &       &       &       &       & &       &      &      &       & \\
\hline
MSE              &          & &       &       &  2.66 &  0.22 & -2.44 & &       &      & 0.86 & -0.05 & -0.91 \\
MAE              &          & &       &       &  2.78 &  1.28 &  2.54 & &       &      & 0.86 &  1.12 & 0.94 \\
MARE             &          & &       &       &71.47\%&31.23\%&72.56\%& &       &      &16.63\%&30.06\%& 32.43\% \\
\hline
       &            &        & \multicolumn{5}{c}{PBE} &        & \multicolumn{5}{c}{PBE0} \\
\cline{4-8}\cline{10-14}
NF$_3$-HCN       & 1.67$^a$ & &  1.42 &  1.83 & -0.25 &  0.16 &  0.41 & &  1.48 &  1.91 & -0.19 &  0.24 &  0.43 \\
C$_2$H$_4$-F$_2$ & 1.69$^b$ & &  5.39 &  1.11 &  3.70 & -0.58 & -4.28 & &  3.05 &  1.13 &  1.36 & -0.56 & -1.92 \\
NF$_3$-HNC       & 2.31$^a$ & &  3.49 &  3.61 &  1.18 &  1.30 &  0.12 & &  2.97 &  3.48 &  0.66 &  1.17 &  0.51 \\
C$_2$H$_4$-Cl$_2$& 2.60$^c$ & &  7.61 &  6.09 &  5.06 &  3.54 & -1.52 & &  5.96 &  6.38 &  3.41 &  3.83 &  0.42 \\ 
NH$_3$-F$_2$     & 2.88$^b$ & &  8.73 &  1.83 &  5.85 & -1.05 & -6.90 & &  4.87 &  1.89 &  1.99 & -0.99 & -2.98 \\
NF$_3$-ClF       & 2.92$^a$ & &  5.01 &  2.87 &  2.09 & -0.05 & -2.14 & &  3.59 &  2.77 &  0.67 & -0.15 & -0.82 \\
NF$_3$-HF        & 2.92$^a$ & &  4.73 &  3.82 &  1.81 &  0.90 & -0.91 & &  4.02 &  3.80 &  1.10 &  0.88 & -0.22 \\
C$_2$H$_2$-ClF   & 6.07$^b$ & & 10.56 &  6.85 &  4.49 &  0.78 & -3.71 & &  8.84 &  7.33 &  2.77 &  1.26 & -1.51 \\ 
HCN-ClF          & 7.74$^b$ & &  9.63 &  8.01 &  1.89 &  0.27 & -1.62 & &  8.46 &  8.43 &  0.72 &  0.69 & -0.03 \\
NH$_3$-Cl$_2$    & 7.78$^b$ & & 13.48 & 10.64 &  5.70 &  2.86 & -2.84 & & 10.90 & 10.69 &  3.12 &  2.91 & -0.21 \\
H$_2$O-ClF       & 8.54$^b$ & & 11.63 &  9.21 &  3.09 &  0.67 & -2.42 & & 10.21 &  9.76 &  1.67 &  1.22 & -0.45 \\
NH$_3$-ClF       & 16.92$^b$& & 28.84 & 24.40 & 11.92 &  7.48 & -4.44 & & 25.62 & 24.31 &  8.70 &  7.39 & -1.31 \\
%                 &          & &       &       &       &       &        & &       &      &       &       & \\
\hline
MSE              &          & &       &       &  3.88 &  1.36 & -2.52  & &       &      &  2.17 &  1.49 & -0.67 \\
MAE              &          & &       &       &  3.92 &  1.64 &  2.61  & &       &      &  2.20 &  1.77 &  0.90 \\
MARE             &          & &       &       &91.53\%&34.43\%& 71.65\%& &       &      &45.82\%&36.92\%& 29.83\% \\
\hline\hline
\end{tabular}
\begin{flushleft}
$^a$ CCSD(T)-F12b/VTZ-F12, Ref. ~\citenum{steinmann12}.  \\
$^b$ W1 benchmark calculations, Ref. ~\citenum{truhlar05}.   \\
$^c$ CCSD(T) extrapolated to CBS limit, this work.
\end{flushleft}
%\end{scriptsize}
\end{table*}

A first inspection of the results, focusing the attention on $E_b^\gks$
and $\Delta^\gks$, confirms the finding,
well established in literature \cite{prissette78,
steinmann12,ruiz96,garcia2000,karpfen03,liao03}, that hybrid functionals
outperform semilocal ones for the calculation of interaction energies
of ground-state charge-transfer complexes. 
In fact, BHLYP with a MARE of 16.63\% is the most accurate method. However,
when interaction energies from embedding calculations are considered
(i.e. $E_b^{\text{emb}}$), all methods yield rather similar performances with 
MAREs in the range 30-37\% (for a discussion of the issues related to the 
comparison of embedding binding energies with experimental or 
theoretical references see e.g. Refs. 
\cite{wesolowski02inter,kevorkyants06hyd,dulak07}). 
This suggests that a strong error
cancellation occurs for embedding calculations using semilocal functionals,
while the cancellation is much less in the case of hybrid approaches.
The reason for this behavior can be traced back to the fact that,
in general, conventional semilocal functionals overestimate 
significantly the binding energy of charge-transfer complexes
\cite{ruiz96,garcia97,garcia2000,karpfen03,liao03,2truhlar05,steinmann12}
(MSE is largely positive) while, on the other hand, when used in methods
based on the FDE theory they show usually a tendency to produce 
too low interaction energies ($\Delta E$ is 
negative in most cases). On the contrary,
hybrid functionals are quite accurate for the calculation of the
interaction energies of charge-transfer complexes and, at the same
time, corresponding embedding calculations reproduce with good
accuracy the parent supermolecular GKS calculations.

Concerning the error on embedding energies ($\Delta E$), we note
that the methods using hybrid XC functionals definitely outperform 
the ones employing semilocal XC approximations, yielding average deviations
more than twice smaller. This improvement can be partially reconduced,
in analogy to what shown in Refs. ~\citenum{laricchia-lhf2011,larihyben}, 
to a smaller overlap of the subsystem densities (i.e. to a smaller SIE) at the
hybrid level, that leads to a reduction of the error
on embedding energy. This effect is best recognized if we consider a set of
systems where only one subsystem is varying, as for NH$_3$
interacting with Cl$_2$, ClF, and F$_2$ respectively.
In this case the value of $\Delta E$ is increasing from 2.6 to about
7 mHa when the Cl atoms are replaced by F atoms and this increase nicely
correlates to the fact that the fluorine atom is affected by a larger 
self-interaction error than chlorine, so that it is plagued by a stronger 
overestimation of the diffuseness of the electron density at semilocal 
level of theory. 
However, in general 
it is not possible to find a trend valid for all the systems due to the  
different details of the interaction when both subsystems are changed.

In fact, despite the inclusion of exact exchange in the calculations may help 
to reduce the errors on embedding energies by lowering the effective overlap
between the two subsystems, this is not the only effect determining the 
accuracy of the GKS-FDE energy calculations. In particular for hybrid methods
still a subtle error compensation between kinetic, relaxation, and 
exchange contributions occurs. Thus, to shed light on the different aspects 
determining the embedding energy error we can recur to the embedding energy 
decomposition introduced in Ref. ~\citenum{larihyben}.
In Tab. \ref{tab:errdecom} we report the energy error
contributions $\Delta T+\Delta D$ and $\Delta X^S$ for all the 
hybrid methods (BHLYP and PBE0) and system considered. For each
energy error contribution we report the MSE, MAE and MARE. 
The contributions due to kinetic ($\Delta T$) and relaxation ($\Delta D$)
effects are reported summed together because both terms
yield very large values (100-200\% of the binding energies) 
and with opposite sign, so they only contribute to the total error 
through a strong error cancellation.
\begin{table}[h!]
\caption{\label{tab:errdecom} Kinetic-relaxation
($\Delta T+\Delta D$), and exchange ($\Delta X^S$) contributions to the
embedding energy error (see Ref. ~\citenum{larihyben} for details).
In all calculations we used the revAPBEk functional for the non-additive 
kinetic energies and a supermolecular def2-TZVPPD basis set. 
At the bottom
of each panel the mean signed error (MSE), mean absolute error (MAE) and
the mean absolute relative error (MARE) are reported. For the 
$\Delta T+\Delta D$ values also the semilocal differential relative error 
(SDRE; see text) is reported.  For $\Delta X^S$ we computed
the exchange differential error (XDE) and the exchange differential 
relative error (XDRE) \cite{larihyben}. All values are in mHa.}
%\begin{scriptsize}
\begin{ruledtabular}
\begin{tabular} {lcc}
%\hline\hline
Systems      & BHLYP &  PBE0  \\
\hline
 & \multicolumn{2}{c}{$\Delta T+\Delta D$} \\
\cline{2-3} 
%\\
%#-----------------------------------------------------------------------
%#          SYSTEMS             BHLYP       PBE0
%#                              DeltaT + DeltaD
%#-----------------------------------------------------------------------
            NF$_3$-HCN &      -0.4 &      -0.4 \\ 
      C$_2$H$_4$-F$_2$ &       2.6 &       3.4 \\ 
            NF$_3$-HNC &      -0.3 &      -0.2 \\ 
     C$_2$H$_4$-Cl$_2$ &       0.3 &       1.1 \\ 
          NH$_3$-F$_2$ &       3.7 &       5.5 \\ 
            NF$_3$-ClF &       1.1 &       1.8 \\ 
             NF$_3$-HF &       0.6 &       0.8 \\ 
        C$_2$H$_2$-ClF &       2.2 &       3.3 \\ 
               HCN-ClF &       0.4 &       1.2 \\ 
         NH$_3$-Cl$_2$ &       0.9 &       2.3 \\ 
            H$_2$O-ClF &       0.8 &       1.9 \\ 
            NH$_3$-ClF &       2.9 &       4.3 \\ 
%\\
\hline
MAE                    &       1.35 &      2.19 \\
MSE                    &       1.23 &      2.09 \\
MARE                   &      39.19\% &   58.65\% \\
SDRE                   &     -33.51\% &  -12.97\% \\
%#-----------------------------------------------------------------------
%#                                  DeltaX^S
%#-----------------------------------------------------------------------
 & \multicolumn{2}{c}{$\Delta X^S$} \\
\cline{2-3}
            NF$_3$-HCN &       1.0 &      -0.0 \\ 
      C$_2$H$_4$-F$_2$ &      -0.4 &      -1.5 \\ 
            NF$_3$-HNC &       0.5 &      -0.3 \\ 
     C$_2$H$_4$-Cl$_2$ &      -0.4 &      -1.6 \\ 
          NH$_3$-F$_2$ &      -1.2 &      -2.5 \\ 
            NF$_3$-ClF &       0.4 &      -0.9 \\ 
             NF$_3$-HF &       0.2 &      -0.6 \\ 
        C$_2$H$_2$-ClF &      -0.5 &      -1.7 \\ 
               HCN-ClF &       0.3 &      -1.2 \\ 
         NH$_3$-Cl$_2$ &      -1.0 &      -2.1 \\ 
            H$_2$O-ClF &       0.0 &      -1.5 \\ 
            NH$_3$-ClF &      -2.7 &      -3.0 \\ 
%\\
\hline
MAE                    &       0.72 &      1.41 \\
MSE                    &      -0.32 &	   -1.41 \\
MARE                   &      19.10\% &	   34.10\% \\
%#-----------------------------------------------------------------------
%\\
\hline
XDE                    &      -0.41 &	   -1.28 \\
XDRE                   &      -6.79\% &	  -28.83\% \\
%\hline\hline
\end{tabular}
\end{ruledtabular}
%\end{scriptsize}
\end{table}

The analysis of the $\Delta T+\Delta D$ contributions shows that
the use of the subsystem electron densities from hybrid calculations can
provide a significant improvement of the semilocal embedding energies
(the MAEs are 1.35 and 2.19 to be compared with 2.54 and 2.61), although
the errors are still larger than the global ones 
reported in Tab. \ref{tab:enerr} (MAEs of 0.94 and 0.90 for BHLYP and PBE0,
respectively). This improvement is determined exclusively
from the reduced overlap displayed by the densities obtained from hybrid
calculations, since here no exchange approximation is considered.
To make a quantitative evaluation of the effect it is possible to 
use the semilocal differential relative error (SDRE) statistical 
indicator, defined as
\begin{equation}
\text{SDRE}= \frac{1}{N}\sum_{i=1}^N\frac{\abs{\Delta T_i+\Delta D_i}
-\abs{\Delta E_i^\GGA}}{E_{b,i}^{ref}} ,
\end{equation}
with the sum extend over all the $N$ systems reported in the table.
A negative value of SDRE indicates that using subsystem densities from
a given hybrid calculation improves on average the calculation of kinetic and 
relaxation energy contributions over the use of GGA densities.
Using the SDRE indicator it is thus possible to compare energetic error 
terms with the same explicit level of approximation, i.e. the 
semilocal kinetic energy approximation and the density relaxation error
\cite{larihyben}. Looking at the values of SDRE for the two hybrids considered
in the table, it can be seen that in both cases these methods 
outperform the corresponding semilocal ones (negative values of SDRE). 
This result reflects the good quality of the electron densities computed with
hybrid embedding methods (see next section) 
and evidences once more the importance of the density overlap in this context.
In fact, the larger improvement is obtained for BHLYP (SDRE of -33.51\%) 
due to its higher content of exact exchange with respect to 
PBE0 (SDRE of -12.97\%).

Considering the exchange contributions to the embedding error ($\Delta X^S$)
we note immediately that, for both methods, they are in general of opposite
sign and of the same order of magnitude as the $\Delta T+\Delta D$ terms.
Hence, an important error cancellation effect occurs in all cases. This effect
can be measured by considering the exchange differential error (XDE) and 
the exchange differential relative error (XDRE) 
statistical indicators \cite{larihyben}
\begin{eqnarray}
\text{XDE}&=& \frac{1}{N}\sum_{i=1}^N\abs{\Delta E_i}
-\abs{\Delta T_i+\Delta D_i} , \\
\text{XDRE}&=& \frac{1}{N}\sum_{i=1}^N\frac{\abs{\Delta E_i}
-\abs{\Delta T_i+\Delta D_i}}{E_{b,i}^{ref}} \ ,
\end{eqnarray}
which provide (by increasingly negative values) an estimation of the
improvement due to the error cancellation effect induced by the additional 
approximation on the exchange nonadditive embedding terms.
For BHLYP and PBE0 the XDE and XDRE values are negative, indicating, as expected, 
an error cancellation between the kinetic-relaxation and the exchange 
contributions. However, the XDE is three times larger for PBE0 than for BHLYP, so
in the former case a much more effective error cancellation occurs.
This explains why PBE0 has the smaller energy error 
$\Delta E$ in Tab. \ref{tab:enerr}, despite it
has kinetic-relaxation errors significantly larger than BHLYP.

\subsection{Embedding density}
\label{sect_dens}
\begin{figure}   
\includegraphics[width=\columnwidth]{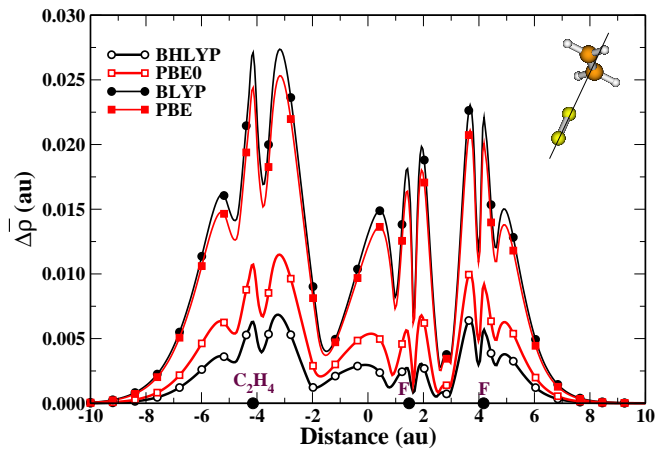}
\caption{\label{denserr} Absolute deviation from the reference supermolecular 
electron density of plane-averaged embedding densities obtained from 
different embedding approaches, for the C$_2$H$_4$-F$_2$ complex. 
The filled circles on the $x$-axis denote the atoms' positions.}
\end{figure}

In this section we consider the ability of different embedding methods
to reproduce the electron density of supermolecular charge-transfer systems. 
This is an important test for embedding approaches and provides direct insight
into the quality of the embedding potential \cite{laricchia2010,laricchia-lhf2011,
laricchia2011,jacobviss08,kiew08,govind09,fux10,beyhan10}.
To discuss this issue we report in Fig. \ref{denserr} the plot
of the absolute deviation of the plane-averaged densities ($\Delta\bar{\rho}$; 
see Eq. (\ref{drho})) for the C$_2$H$_4$-F$_2$ complex, as an example.
For other systems similar plots are obtained (not reported).

The figure shows that, unlike for hydrogen bond systems
\cite{laricchia2010,laricchia-lhf2011}, which are 
prototypical examples for FDE-based studies, the density errors are quite
different for different methods, especially if hybrid and GGA approaches are
compared.
This trend can be well captured by considering the
integrated errors on embedding densities 
($\xi$; see Eq. (\ref{eq:xi})), as resulting from different 
methods, which are reported in Tab. \ref{tab:dens_err}.
\begin{table}[h!]
\caption{\label{tab:dens_err} Valence density errors $\xi$ for 
GKS-FDE calculations (BHLYP and PBE0) and
FDE-based calculations using conventional BLYP and PBE functionals.
In all calculations we used the revAPBEk functional for the nonadditive 
kinetic energies and a supermolecular def2-TZVPPD basis set.
In the last line the mean absolute error (MAE) is reported.}
%\begin{scriptsize}
\begin{ruledtabular}
\begin{tabular} {lcccc}
%\hline\hline
 & \multicolumn{2}{c}{Hybrid} & \multicolumn{2}{c}{Semilocal} \\
Systems      & BHLYP &  PBE0 & BLYP &  PBE \\
\hline
            NF$_3$-HCN &    0.13 &    0.24 &    0.29 &    0.29 \\ 
      C$_2$H$_4$-F$_2$ &    1.58 &    2.75 &    6.94 &    6.35 \\ 
            NF$_3$-HNC &    0.41 &    0.49 &    0.58 &    0.58 \\ 
     C$_2$H$_4$-Cl$_2$ &    3.34 &    4.31 &    5.83 &    5.77 \\ 
          NH$_3$-F$_2$ &    2.24 &    4.36 &    9.84 &    9.59 \\ 
            NF$_3$-ClF &    0.64 &    0.99 &    1.64 &    1.73 \\ 
             NF$_3$-HF &    0.60 &    0.74 &    0.96 &    0.94 \\ 
        C$_2$H$_2$-ClF &    3.14 &    4.32 &    6.03 &    6.02 \\ 
               HCN-ClF &    1.76 &    2.33 &    3.13 &    3.21 \\ 
         NH$_3$-Cl$_2$ &    3.98 &    5.48 &    7.45 &    7.60 \\ 
            H$_2$O-ClF &    2.37 &    3.41 &    4.89 &    5.06 \\
            NH$_3$-ClF &    8.07 &    9.37 &   11.27 &   11.15 \\
\\
%#----------------------------------------------------------------
MAE                    &    2.35 &    3.23 &    4.90 &    4.86 \\
%\hline\hline
\end{tabular}
\end{ruledtabular}
%\end{scriptsize}
\end{table}
These data, along with the proportionality 
of the $\Delta\bar{\rho}(z)$ profiles reported in Fig. \ref{denserr}
(the embedding errors in the density have a similar spatial 
distribution),
show that for all systems the errors on embedded densities
are directly correlated to the amount of exact exchange included in
the calculations (see also next section). 
In fact, the best results are found for BHLYP, which
includes 50\% of Hartree-Fock exchange and has a MAE of only 2.35, comparable
to that found in the case of hydrogen-bond and dipole-dipole 
interactions \cite{laricchia2010,laricchia-lhf2011}. 
On the contrary, methods based on GGA XC functionals yield 
significantly larger errors, more than twice as large as the BHLYP ones.
We note that such large variations of the embedding density error with the
method are rather unusual and are not encountered in the case of
hydrogen bond and dipole-dipole 
complexes \cite{laricchia2010,laricchia-lhf2011}. This indicates the special
relevance of the inclusion of exact exchange for the treatment of charge-transfer
complexes via FDE-based methods and fully agrees with the analysis made in the
previous section about the role of relaxation effects for the energy errors
of different methods.

Beside the total quality of the embedding density, in the present context
another quantity of great interest is the net charge transfer predicted by
different methods for the complexes under examination.
The results for the various approaches, together with reference MP2 values 
are reported in Tab. \ref{tab:ct}.
\begin{table}
\caption{\label{tab:ct} Charge transfer $\chi$
resulting from conventional supermolecular DFT and embedding calculations
for several test charge-transfer complexes.
The mean absolute error (MAE), the mean signed 
error (MSE), and the mean absolute relative error (MARE) respect to the
reference MP2 charge transfer are also reported$^a$.}
%\begin{scriptsize}
\begin{ruledtabular}
\begin{tabular} {lrccrccrc}
%\hline\hline
Systems  & $\;$ & $\chi^{\KS}$ & $\chi^{\text{emb}}$ & $\;$ & 
$\chi^{\gks}$ & $\chi^{\text{emb}}$ & $\;$ & MP2 \\
\hline
         &      & \multicolumn{2}{c}{BLYP}   &     & \multicolumn{2}{c}{BHLYP}  & & \\
\cline{3-4}\cline{6-7}
NF$_3$-HCN & & 0.017 & 0.011 & & 0.012 & 0.010 & & 0.009 \\
C$_2$H$_4$-F$_2$ & & 0.082 & 0.014 & & 0.029 & 0.015 & & 0.018 \\
NF$_3$-HNC & & 0.007 & 0.007 & & 0.006 & 0.006 & & 0.005 \\
C$_2$H$_4$-Cl$_2$ & & 0.093 & 0.044 & & 0.061 & 0.045 & & 0.053 \\
NH$_3$-F$_2$ & & 0.105 & 0.018 & & 0.032 & 0.015 & & 0.026 \\
NF$_3$-ClF & & 0.035 & 0.013 & & 0.015 & 0.009 & & 0.013 \\
NF$_3$-HF & & 0.025 & 0.013 & & 0.018 & 0.012 & & 0.018 \\
C$_2$H$_2$-ClF & & 0.102 & 0.051 & & 0.067 & 0.049 & & 0.052 \\
HCN-ClF & & 0.050 & 0.028 & & 0.033 & 0.025 & & 0.034 \\
NH$_3$-Cl$_2$ & & 0.122 & 0.060 & & 0.064 & 0.044 & & 0.061 \\
H$_2$O-ClF & & 0.072 & 0.034 & & 0.041 & 0.028 & & 0.042 \\
NH$_3$-ClF & & 0.217 & 0.209 & & 0.163 & 0.126 & & 0.177 \\
          & &       &       & &       &       & & \\
MSE	  & & 0.035 & -0.001 & & 0.003 & -0.011 & & \\ 	
MAE	  & & 0.035 & 0.007 & &	0.005 &	0.011& & \\	
MARE	  & & 118.07\% & 18.64\% & & 17.50\% & 25.63\% & & \\
\hline
         &      & \multicolumn{2}{c}{PBE}   &     & \multicolumn{2}{c}{PBE0}  & & \\
\cline{3-4}\cline{6-7}
NF$_3$-HCN & & 0.016 & 0.011 & & 0.013 & 0.011 & & 0.009 \\
C$_2$H$_4$-F$_2$ & & 0.074 & 0.013 & & 0.039 & 0.013 & & 0.018 \\
NF$_3$-HNC & & 0.007 & 0.007 & & 0.006 & 0.006 & & 0.005 \\
C$_2$H$_4$-Cl$_2$ & & 0.093 & 0.044 & & 0.074 & 0.044 & & 0.053 \\
NH$_3$-F$_2$ & & 0.101 & 0.017 & & 0.052 & 0.015 & & 0.026 \\
NF$_3$-ClF & & 0.037 & 0.013 & & 0.024 & 0.011 & & 0.013 \\
NF$_3$-HF & & 0.025 & 0.013 & & 0.020 & 0.013 & & 0.018 \\
C$_2$H$_2$-ClF & & 0.101 & 0.050 & & 0.081 & 0.049 & & 0.052 \\
HCN-ClF & & 0.051 & 0.028 & & 0.040 & 0.027 & & 0.034 \\
NH$_3$-Cl$_2$ & & 0.125 & 0.061 & & 0.089 & 0.050 & & 0.061 \\
H$_2$O-ClF & & 0.074 & 0.035 & & 0.054 & 0.030 & & 0.042 \\
NH$_3$-ClF & & 0.221 & 0.206 & & 0.191 & 0.162 & & 0.177 \\
          & &       &       & &       &       & & \\
MSE	  & & 0.035 & -0.001 & &	0.015 &	-0.006 & & \\
MAE	  & & 0.035 & 0.006 & &	0.015 &  0.007 & & \\
MARE	  & & 113.80\% & 19.26\% & & 48.22\% & 21.00\% & & \\
%\hline\hline
\end{tabular}
\end{ruledtabular}
\begin{flushleft}
$^a$ The statistical indicators are computed using the error
$\Delta\chi^\ell=\chi^\ell-\chi^{\text{MP2}}$ with $\ell=$KS,GKS,emb.
\end{flushleft}
%\end{scriptsize}
\end{table}

The table shows the well-known tendency of the GGA XC functionals
to overestimate the charge transfer \cite{ruiz95,ruiz96},
so that BLYP and PBE yield for $\chi^{\gks}$ MAREs well above 100\%, while
much better results (MARE of 17.5\%) are obtained by including a large fraction
of exact exchange, as in BHLYP. However, a different situation is observed when
embedding calculations are considered. These have, in fact, a marked tendency 
towards an underestimation of the charge transfer, compared to the corresponding
supermolecular calculation on the whole system, probably related to the
excessive repulsive character of the approximated embedding potential.
This behavior is also much more pronounced for GGA methods than for hybrids.
As a consequence, the BHLYP results are slightly worsened compared to 
the reference MP2 ones when embedding is used, while the semilocal methods 
display a strong improvement. 

In order to shine light on this finding, we consider that
embedding densities computed at the semilocal level of theory, when used
to compute the energy, provide higher energies for the complexes 
with respect to corresponding densities from supermolecular
calculations
(semilocal functionals stabilize the energy of charge-transfer complexes by
overestimating the charge transfer to compensate for the absence of long-range
interactions \cite{steinmann12}). 
Because, on the other hand, the isolated systems are not affected by 
this problem, this leads in general to a reduction of the computed binding energy
(see Tab. \ref{tab:enerr}), thus effectively compensating for the tendency of 
the semilocal XC functionals to over-bind. This explains the fact that in Tab.
\ref{tab:enerr}, for semilocal functionals, $\Delta^{\text{emb}}$ has a much
smaller MAE and MARE than $\Delta^\gks$. It is worth noting that the same
considerations also apply to the hybrid approaches, but in these cases the 
effect is smaller. In fact, PBE0 binding energies are slightly improved 
(on average) when computed via an embedding approach, while BHLYP results
end up to suffer from a slight underbinding.

\subsection{Role of the amount of exact exchange}
\label{sect_alpha}
In previous sections we evidenced the important role of the exact exchange
for the description of embedding energies and densities, by comparing
the performance of FDE-based methods using semilocal XC functionals with that of 
GKS-FDE approaches using hybrid functionals based on exactly the same semilocal 
approximation. To push this investigation one step further here we consider more
generally the family of one-parameter hybrid functionals
\begin{figure*}   
\includegraphics[width=\textwidth]{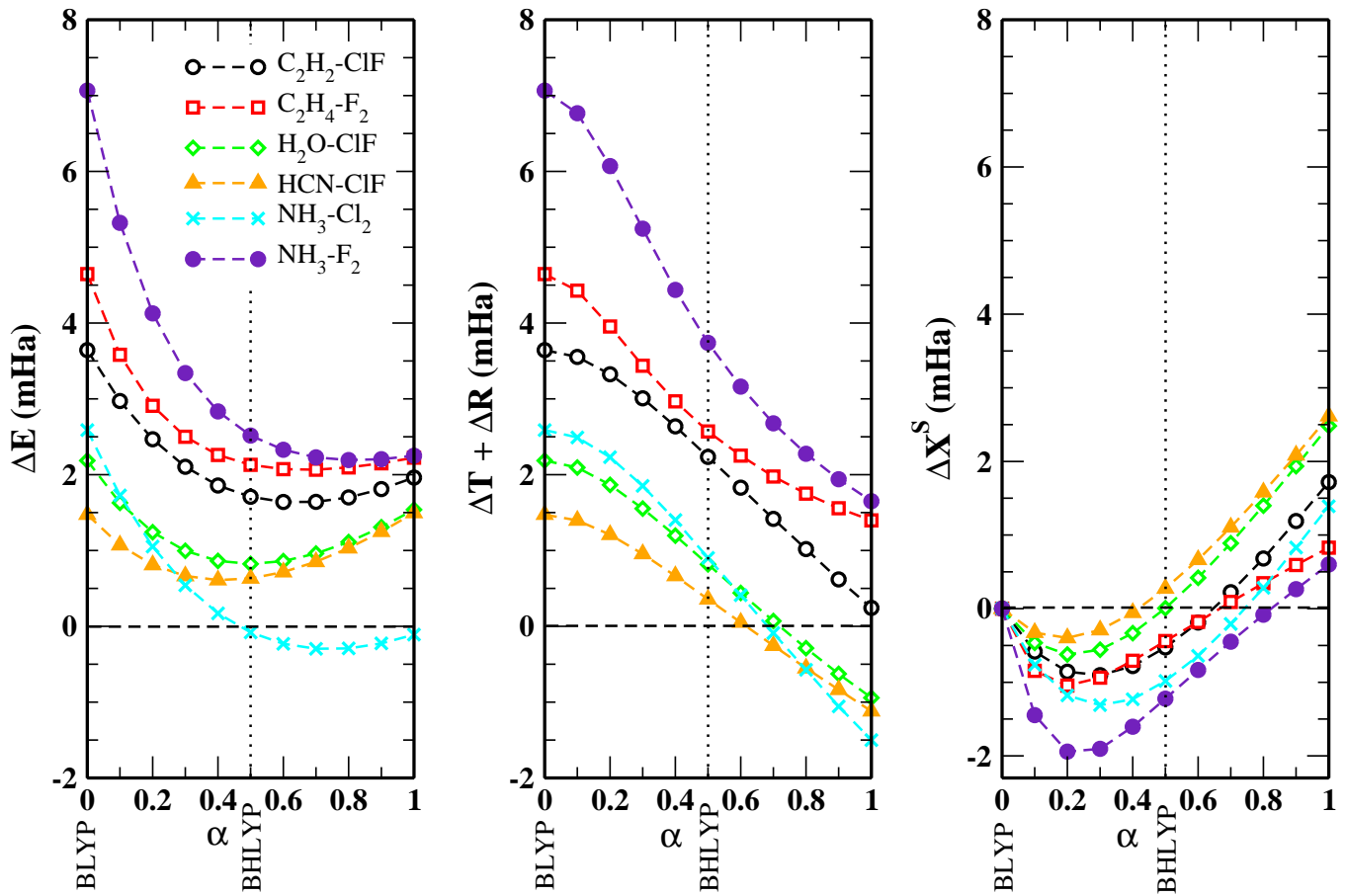}
\caption{\label{fig:alpha} Errors on the embedding energy for selected systems
are reported in the left panel as computed with the family of functionals
of Eq. (\ref{eq:family}). The corresponding contributions $\Delta T+\Delta D$
and $\Delta X^S$ are plotted in the middle and in the right panels,
respectively. In all calculations we used the revAPBEk functional for the 
non-additive kinetic energies and a supermolecular def2-TZVPPD basis set.}
\end{figure*}
\begin{equation}\label{eq:family}
E_\xc^{\text{hyb}}(\alpha)=\alpha E_{HF} + (1-\alpha)E_\xonly^{\text{B88x}}
+ E_\conly^{\text{LYP}}
\end{equation}
where $E_{HF}$ is the Hartree-Fock exchange, $\alpha$ is a parameter, and 
$E_\conly^{\text{LYP}}$ is the Lee-Yang-Parr correlation energy functional
\cite{lyp}. Eq. (\ref{eq:family}) reduces to BLYP for $\alpha=0$,
to BHLYP for $\alpha=0.5$, and was already used to study the role
of exact exchange in embedding calculations of hydrogen-bond and 
dipole-dipole interacting systems in Ref. ~\citenum{larihyben}.

Fig. \ref{fig:alpha} reports, as a function of $\alpha$, 
the errors on the embedding energies
for selected systems as well as the individual 
contributions $\Delta T  + \Delta D$ and $\Delta X^S$.
For most of the systems the error $\Delta E$ is found to decrease with 
$\alpha$ until $\alpha\sim0.5$, and then to slightly increase for
larger values of the parameter. This behavior is a consequence of an error 
cancellation between the $\Delta T  + \Delta D$ contribution (middle panel), 
which starts from a rather large positive value and decreases with $\alpha$, 
and the $\Delta X^S$ term (right panel), which shows instead an evident 
parabolic shape with minima located around $\alpha=0.3$ and provides mostly
negative contributions for $\alpha<0.5$.
This behavior is quite different from the one observed for 
dipole-dipole and hydrogen-bond interactions (see Figure I of 
Ref. ~\citenum{larihyben}), which is generally characterized by positive
contributions of $\Delta X^S$, increasing with $\alpha$.
The present finding thus shows that for charge-transfer systems a more
delicate balancing must be found between the need to 
increase the exact-exchange 
contribution, in order to reduce the density overlap, and the necessity to
obtain the correct description of long-range XC interactions in the 
hybrid functional. 

\begin{figure}
\includegraphics[width=\columnwidth]{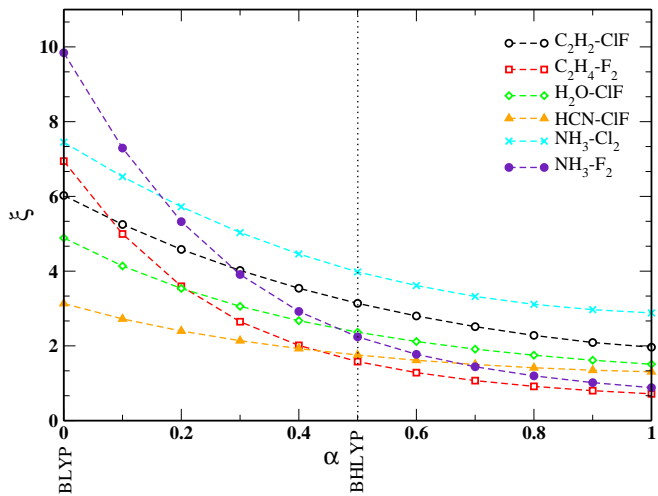}
\caption{\label{adens} Errors on the embedding density for selected systems 
as computed with the family of functionals of Eq. (\ref{eq:family}). 
In all calculations we used the revAPBEk functional for the non-additive 
kinetic energies and a supermolecular def2-TZVPPD basis set.}
\end{figure}
The family of one-parameter hybrid functionals can be also be employed 
to study the effect of the exact exchange on the description of
embedded densities. The result of this study was already partially anticipated
in previous sections and it is clearly shown in Fig. \ref{adens}:
by increasing the amount of exact exchange the error on the embedding density
is monotonically decreased. As mentioned in previous discussions and in 
Ref. ~\citenum{laricchia-lhf2011} this behavior 
is mainly due to the reduction of the overlap
of the subsystem densities consequent to the reduction of the SIE.
Of course, the effect is somehow attenuated for very large values of the
parameter $\alpha$ because of the increased importance of the semilocal 
approximation used for the exchange term in the GKS-FDE scheme in these cases, 
as shown in right panel of Fig. \ref{fig:alpha}.

\begin{figure} 
\includegraphics[width=\columnwidth]{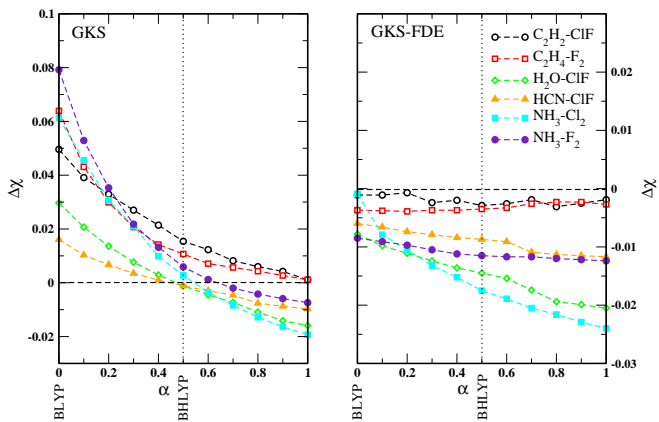}
\caption{\label{ct_alpha} Difference between the charge transfer computed 
with functionals defined by Eq. (\ref{eq:family}) and the reference MP2 value. 
In the left panel we report the results of GKS calculations on whole systems; 
in the right panel the corresponding embedding results are shown. 
In all calculations we used the revAPBEk functional for the non-additive 
kinetic energies and a supermolecular def2-TZVPPD basis set.}
\end{figure}

Finally, we can consider the ability of methods including different amounts
of exact exchange to correctly reproduce the charge transfer. We thus
report in Fig. \ref{ct_alpha} the difference $\Delta \chi$ between the charge
transfer computed for each value of $\alpha$, with or without the embedding
treatment, and the reference values obtained from MP2 relaxed densities.
The plot shows that considering full GKS calculations the charge 
transfer is generally overestimated at the GGA level and better
agreement with the reference is only found when a significative amount of 
Hartree-Fock exchange is considered. 
On the other hand, for embedding calculations the GKS charge transfer
can be only reproduced when exact exchange is included, while
calculations using a semilocal XC potential yield a strong
underestimation with respect to the corresponding supermolecular calculation.
In fact, the charge transfer computed from embedding methods 
appears to be almost independent from the percentage of exact 
exchange included in the XC functional.
To understand these results we must consider that within FDE-based methods
the number of electrons on each subsystem is fixed a priori, so that no
real charge transfer between them can occur. The charge transfer must be 
instead mimicked by a strong polarization of the two subsystem densities,
in such a way that their sum will give the same spatial distribution as the 
electron density of the full system.
The role of the embedding potential is thus to provide a polarization of 
the subsystems, while each subsystem must respond properly to this 
perturbation. In the case of semilocal functionals they are well known to 
provide a poor description of polarization \cite{truhlar06}, thus a poor
description of charge transfer must be expected. On the contrary,
hybrid functionals have been shown to give a good response to polarization 
perturbations within the GKS-FDE scheme \cite{laricchia2010}, thus they can
provide a good reproduction of the charge transfer between the
subsystems. 

\subsection{Role of charge transfer}
\label{sec:ct}
In an analysis of the performance of embedding methods for the simulation
of ground-state charge-transfer complexes it is obviously interesting
to study in details what is the relationship between the outcome of
the calculations and the magnitude of the charge transfer. However, this
study cannot be performed straightforwardly by a simple comparison of 
different systems displaying different values of the charge transfer
because in this way also many other details of the interaction are changed
at the same time.
For this reason in this section we prefer to focus the attention on a single 
system, namely the linear HCN-ClF complex, and force a modulation of the
charge transfer through the application of an external constant electric
field directed along the axis of the complex.

\begin{figure}   
\includegraphics[width=\columnwidth]{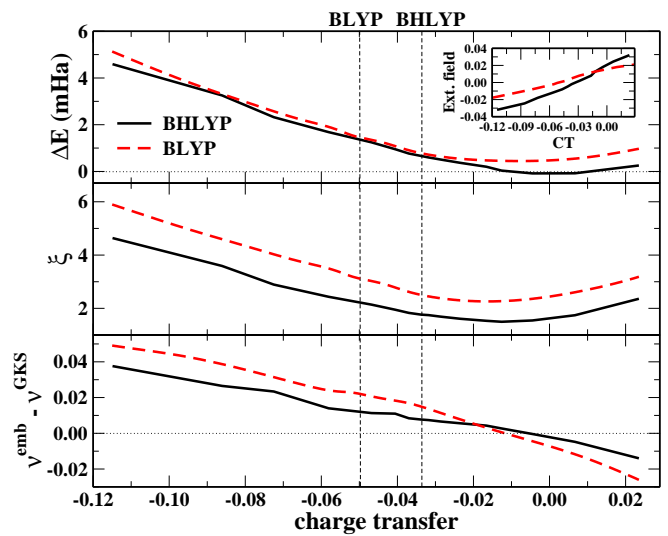}
\caption{\label{field} Errors on embedding energy ($\Delta E$), embedding 
density ($\xi$), and the computed signed charge transfer 
on the dihalogen monomer ClF ($\nu^{\text{emb}}-\nu^{\gks}$) 
as a function of the charge transfer induced 
(at the GKS level) by the applied external field, for BHLYP and BLYP.
The inset reports the external electric field used to generate different 
charge transfer values.}
\end{figure}

In Fig. \ref{field} we report the error on embedding energy ($\Delta E$),
the error on embedding density ($\xi$), and the error on the
computed net charge on the dihalogen monomer ClF 
($\nu^{\text{emb}}-\nu^{\gks}$) as a function of the
charge transfer induced (at the GKS level) by the applied external field
for the two functionals BHLYP and BLYP. The corresponding electric field
is reported in the inset of Fig. \ref{field}.
The figure shows that both functionals display the same behavior
with the induced charge transfer, with all errors increasing as the charge 
transfer grows. In fact, the minima of the embedding errors are approximately
located at zero charge transfer.
The energy error decomposition (not reported) confirms that the increase 
of the energy error is essentially due to a large increase 
in the kinetic and relaxation contributions.

Moreover, the lowest panel of Fig. \ref{field} shows, in agreement
with our analysis of previous sections,
that the increase of charge transfer is in linear correspondence with 
an increase of the inaccuracy in its evaluation by embedding
methods. This implies, as also shown by a comparison of the two upper panels of
Fig. \ref{field} with the lower one, that the embedding errors 
are correlated (almost linearly) with the inaccuracy
of embedding approaches to correctly describe the amount of charge transfer.

\subsection{Role of geometry}
\label{sec:geo}
All previous investigations have been carried out at reference geometries.
However, for charge-transfer complexes large differences
in the equilibrium geometry can be expected for different methods.
In particular, hybrid XC functionals are able to produce
equilibrium structures very close to the reference ones but semilocal
XC approximations lead to significantly shorter 
intermolecular distances \cite{ruiz95,ruiz96,garcia97,karpfen03}. 
As shown in Ref. ~\citenum{larihyben} the geometry and
the intermolecular distance play an important role in determining the 
accuracy of embedding calculations. It is therefore interesting, in the 
present context, to assess this issue for the limits represented by the 
structures optimized at the semilocal BLYP and hybrid BHLYP level.

\begin{figure}   
\includegraphics[width=\columnwidth]{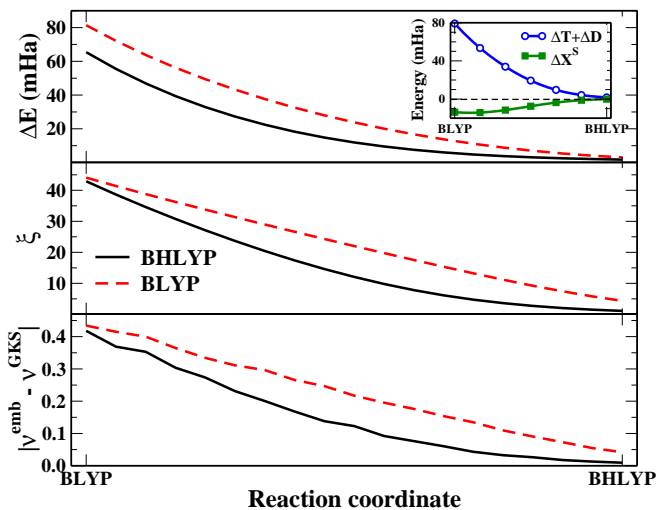}
\caption{\label{geom} Errors on embedding energy ($\Delta E$), embedding 
density ($\xi$), and the computed net charge on the ethylene
monomer ($\nu^{\text{emb}}-\nu^{\gks}$) at different geometries interpolating 
between the BLYP and BHLYP equilibrium geometry of the C$_2$H$_4$-F$_2$ 
complex. In the inset the embedding error decomposition for the 
BHLYP case is reported.}
\end{figure}

To this end we consider in Fig. \ref{geom} the error on 
embedding energy ($\Delta E$), the error on embedding density ($\xi$), 
and the error on the computed monomer charge ($\nu^{\text{emb}}-\nu^{\gks}$) 
at different geometries interpolating linearly between the
BLYP and the BHLYP optimized geometries of the complex C$_2$H$_4$--F$_2$
(similar results are found for other systems).
The intermolecular distance between the F$_2$ monomer 
and the plane of the ethylene is $1.91$ \AA~ and $2.92$ \AA~ for
BLYP and BHLYP, respectively. 
The figure shows the important role of the geometry for all the embedding 
results. This finding can be understood considering that at small 
intermolecular distances, i.e. at the BLYP geometry, 
the overlap between the electron densities of the two subsystems rapidly grows.
Moreover, because at reduced distances between the subsystems 
the orbital interactions are enhanced, also the charge transfer 
is highly increased, together with the error in its evaluation.
As a consequence, large errors can be expected for the embedded densities, due
to the well-known limitations of nonadditive kinetic potentials
for strongly overlapping densities \cite{larihyben,wesobook}.
Similarly, large errors are found for the embedding energies.
In this case an energy decomposition of the BHLYP embedding error 
(see inset of Fig. \ref{geom}) further confirms
that the increase of the error is essentially due to a large increase
of the relaxation-kinetic term, while only a minor role is
played by the exchange term $\Delta X^S$. Nevertheless, the latter
being always opposite in sign to the relaxation-kinetic contribution
provides a moderate error cancellation which improves the final performance
of the hybrid method.

The results of Fig. \ref{geom} indicate in conclusion that,
for the simulation of charge-transfer complexes via embedding approaches,
the semilocal methods may display serious problems related to the use of
relaxed structures (e.g. in geometry optimizations or molecular dynamics),
beside the limitations already highlighted in the previous sections.
Therefore, the hybrid approaches appear definitely more robust in such 
applications.

\subsection{Orbital energies}
\label{sect_orb}

In previous sections we showed that only embedding approaches making use
of hybrid XC functionals can reproduce the ground-state charge transfer 
of the investigated complexes with reasonable accuracy, while a strong
underestimation occurs when semilocal approaches are used.
This fact has important consequences on the accuracy of 
embedding calculations, that nevertheless are partially, or even mostly, 
mitigated by error compensation effects.
The improper description of the charge transfer can have however more subtle
effects, especially on the one-electron spectrum of the subsystems.
This is an important issue, since the molecular orbitals of the
embedded subsystems can be used as the basis for an analysis of the
properties of the interacting subsystems and/or as input quantities in
response calculations \cite{casidaweso,neugetd07}.
\begin{figure}   
\includegraphics[width=\columnwidth]{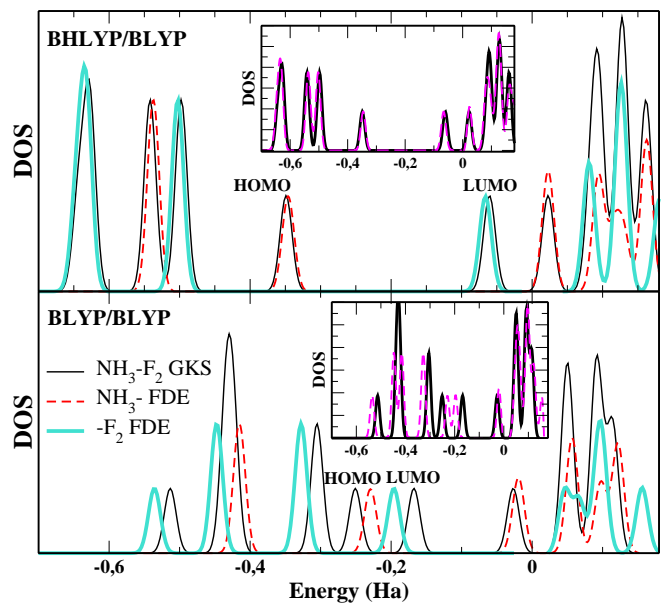}
\caption{\label{fig:dos} Density of states (DOS) for both GKS and 
hybrid BHLYP (upper panel) and the semilocal BLYP (bottom 
panel) embedding calculations on the charge-transfer complex NH$_3$-F$_2$.
The solid black line indicates GKS DOS of the whole system; the red dashed 
and solid turquoise lines indicate DOS from embedding calculations on the 
monomers NH$_3$ and F$_2$ respectively. 
In the insets the GKS DOS (thick black line) is compared with the sum
of the two subsystems DOSs (dashed magenta line).
Energy values are in Ha.}
\end{figure}

The driving force for ground-state charge transfer in molecular
complexes is a sizable difference in the chemical potential between
the constituting (isolated) subsystems, so that,
when the subsystems start to interact, a fraction of the electron density
must be moved from one fragment to another to equilibrate it.
The final chemical potential of the complex is an average between those of the
initial non-interacting subsystems and it is determining the 
offset for the Kohn-Sham potential, thus de facto fixing the energy of
the molecular orbitals.
In an ideal exact FDE-based embedding calculation the embedding potential
shall be able, by definition, to yield for 
the embedded subsystems the same chemical
potential as for the full complex and, of course, shall induce a polarization
able to mimic the exact charge-transfer. As a side consequence, the
single-particle spectrum of the subsystems will be also correct
(as compared to the corresponding supermolecular calculation).
Similar conclusions apply for any approximated embedding calculation having
sufficient accuracy.
On the other hand, if the embedding procedure fails to reproduce the
correct charge transfer (i.e. correct with respect to the calculation
on the full system), the polarization exerted by the embedding potential on
the two subsystems will be insufficient, thus the chemical potential
of the two fragments will be rather different from the one
of the full system (too large in one subsystem and too low in the other).
Consequently, the molecular orbitals of the two subsystems will lay
at too high (for one subsystem) and too low energies (for the other).

The two cases described above fit very well to the embedding calculations
based on hybrid and semilocal functionals, respectively. This is nicely
shown in Fig. \ref{fig:dos} where, as an example, 
we report the density of states (DOS) of the NH$_3$-F$_2$ complex
compared to that of the two embedded subsystems as resulting from 
BHLYP and BLYP embedding calculations.
The comparison is in this case rather fair, since the two fragments
composing the NH$_3$-F$_2$ complex interact only weakly so that the
molecular orbitals of the full system can be easily attributed
to either of the constituent subsystems.
The figure shows very clearly that BHLYP embedding calculations yield
subsystems orbitals which match well the ones from the full 
system (within 0.01 eV).
Instead, the orbitals from the BLYP embedding calculations are systematically
shifted with respect to the ones of the corresponding supermolecular
calculation, with F$_2$ orbitals shifted at lower energies and 
NH$_3$ ones shifted at higher energies, consistently with the 
underestimation of charge transfer from NH$_3$ to F$_2$.
Finally, we recall in favor of hybrid calculations
that even supermolecular calculations using semilocal XC functionals
provide a poor description of the single-particle spectrum, while 
better results can be only obtained by reducing the SIE
\cite{lhf1,LS2}.

\section{Conclusions}
We applied methods based on the frozen density embedding theory to study the
electronic properties of ground-state charge-transfer complexes and
to clarify the ability of such approaches to treat accurately this class
of systems and interactions. This topic is of particular interest due to the
growing popularity of embedding approaches in the field of computational
chemistry and the importance of charge-transfer complexes in many chemical 
applications. 
However, to date, embedding calculations have been mainly presented 
for hydrogen-bond and dipole-dipole interactions, while only few
applications dealt with systems where charge transfer plays a 
significant role \cite{beyhan10,gotz09,dulak06,pavanello1}.

Our work clarifies some of the motivations 
for this lack of embedding calculations
concerning charge-transfer systems, showing that standard embedding 
approaches, making use of semilocal XC functionals, display several
drawbacks in this context. Moreover, despite they can generally yield 
interaction energies of good quality (with respect to accurate references),
this is due to a strong error cancellation effect and does not reflect a
property of the corresponding Kohn-Sham supermolecular calculation.
For these reasons and because semilocal XC functionals are known
to perform rather poorly for charge-transfer complexes even in
conventional Kohn-Sham supermolecular calculations, the use of such
approaches does not appear appropriate.

The introduction of the GKS-FDE method mitigates different
problems in the FDE-based calculations and thus provides a valuable tool
to handle embedding calculations of charge-transfer systems.
In fact, in the present work we showed the relevance of 
exact exchange contributions to reduce the embedding errors and
improve the overall performance of the calculations, through a
reduction of the Coulomb self-interaction error and a better description
of subsystems' electronic properties.

In conclusion, FDE-based approaches making use of hybrid XC functionals
appear as reliable and efficient methods to handle electronic structure
calculations of systems characterized by charge-transfer interactions.
This suggest the possibility of future studies in this direction making
use of these tools to investigate charge-transfer interactions in complex
environments.

\begin{acknowledgments}
This work was partially funded by the European Research Council (ERC) 
Starting Grant FP7 Project DEDOM, Grant No. 207441. The authors thank 
TURBOMOLE GmbH for providing the TURBOMOLE program package and M. 
Margarito for technical support.
\end{acknowledgments}

\end{document}